\documentclass[aps,prl,twocolumn,showpacs,superscriptaddress]{revtex4-1}
\usepackage{graphicx,amsmath,amssymb,amsfonts}
\usepackage[latin1]{inputenc}
\usepackage{array}
\usepackage{multirow}
\usepackage{float}
\usepackage{color}
\usepackage[normalem]{ulem}
\begin{document}

\title{FeGe$_{1-x}$Sb$_x$: a series of novel kagome metals with noncollinear antiferromagnetism}

\author{Jiale Huang}
\author{Chenglin Shang}
\affiliation{Laboratory for Neutron Scattering and Beijing Key Laboratory of Optoelectronic Functional Materials and MicroNano Devices, Department of Physics, Renmin University of China, Beijing 100872, China}
\affiliation{Key Laboratory of Quantum State Construction and Manipulation (Ministry of Education), Renmin University of China, Beijing, 100872, China}

\author{Jianfei Qin}
\affiliation{China Institute of Atomic Energy, PO Box-275-30, Beijing 102413, China}
\author{Feihao Pan}
\author{Bingxian Shi}
\author{Jinchen Wang}
\author{Juanjuan Liu}
\author{Daye Xu}
\author{Hongxia Zhang}
\affiliation{Laboratory for Neutron Scattering and Beijing Key Laboratory of Optoelectronic Functional Materials and MicroNano Devices, Department of Physics, Renmin University of China, Beijing 100872, China}
\affiliation{Key Laboratory of Quantum State Construction and Manipulation (Ministry of Education), Renmin University of China, Beijing, 100872, China}

\author{Hongliang Wang}
\author{Lijie Hao}
\affiliation{China Institute of Atomic Energy, PO Box-275-30, Beijing 102413, China}

\author{Peng Cheng}
\email[Corresponding author: ]{pcheng@ruc.edu.cn}
\affiliation{Laboratory for Neutron Scattering and Beijing Key Laboratory of Optoelectronic Functional Materials and MicroNano Devices, Department of Physics, Renmin University of China, Beijing 100872, China}
\affiliation{Key Laboratory of Quantum State Construction and Manipulation (Ministry of Education), Renmin University of China, Beijing, 100872, China}

\begin{abstract}
Kagome metals are important for exploring emergent phenomena due to the interplay between band topology and electron correlation. Motivated by the recent discovery of charge density wave in a
kagome lattice antiferromagnet FeGe, we investigate the impact of Sb doping on the structural, charge and magnetic order of FeGe. The charge density wave is rapidly suppressed by Sb doping ($\sim$1.5$\%$) and the antiferromagnetic ordering temperature gradually shifts to 280~K for FeGe$_{0.7}$Sb$_{0.3}$. For FeGe$_{1-x}$Sb$_x$ with x$\geqslant0.1$, crystal structures with slightly distorted Fe kagome lattice are formed. Their magnetic anisotropy has significant change, temperature driven spin-reorientation and field-induced spin-flop transitions are identified from magnetization measurement. Interestingly, neutron diffraction reveals noncollinear antiferromagnetic structures widely exists below T$_N$ for all samples with x$\geqslant$0.1. These noncollinear magnetic orders could possibly be unconventional and resulted from onsite repulsion and fillings conditions of kagome flat band, as predicted by a recent thoeretical work.    

\end{abstract}

\maketitle

\section{Introduction}

Kagome lattice hosts peculiar electronic structure with the coexistence of Dirac cones, flat bands and van Hove singularities\cite{Yin2022,Kang2020,Li2021}. In metallic materials with $3d$ transitional metal kagome networks, various novel emergent phenomena including superconductivity, magnetism, anomalous Hall effect and charge order have been observed in recent years\cite{135_SC1,135review,FBFM,FBFM2018,FeSnPRM,TbMn6Sn6,Mn3Sn,CDW1,CDW2}. Therefore, they have become an important platform to explore correlated quantum states intertwined with topological band structures. 

The kagome charge density wave (CDW) has drawn great attentions due to its many-body correlations and topological features\cite{Yin2022}. It was initially discovered in kagome superconductors AV$_3$Sb$_5$ (A=K,Cs,Rb) and found to break time-reversal symmetry with the absence of any long range magnetic order\cite{TRS_Nature2022}. This CDW order is considered to be unconventional, arising from Fermi surface nesting of van Hove singularities and hosting a chiral flux phase which induces anomalous Hall effect\cite{CFP1,CFP2,CFP3,135AHE1,135AHE2}. On the other hand, the magnetism in kagome metals may also be unconventional. It has been proposed that the large density of states from the kagome flat bands could induce ferromagnetism\cite{FBFM,FBFM2018}. However, the coexistence and interplay between CDW and long range magnetic order has not been observed in kagome metals until recently. Hexagonal FeGe with kagome lattice was reported to display a CDW transition at 100~K coupled to the long range antiferromagnetic order below T$_N$=410~K\cite{FeGe}. 

Spectroscopic experiments have reveals an intimate interaction between the CDW order and magnetism in FeGe\cite{FeGe_ARPES,FeGe_STM}. However the origin of this CDW order remains elusive, as well as its relation with anomalous Hall effect and magnetic order. Furthermore, a recent Hartree-Fock analysis shows that unconventional noncollinear antiferromagnetic (AFM) order may exist in the magnetic phase diagram of FeGe tuned by onsite repulsion and flat-band fillings\cite{arxiv2023}. For materials with noncollinear antiferromagnetism, the scalar spin chirality or a nonzero Berry curvature with the spin-orbital coupling may induce strong anisotropic anomalous Hall effect and spin Hall effect\cite{AFM_AHE_T1,AFM_AHE_T2}. These intriguing effects have been realized in Mn$_3$Sn and Mn$_3$Ge with the kagome lattice and received great research interests\cite{Mn3Ge,Mn3Sn,Mn3Sn_spinhall}. Although there are a large number of magnetic kagome metals discovered so far, the noncollinear antiferromagnet seem to be quite rare besides Mn$_3$X (X= Ge, Sn, Ga, Ir) material family\cite{Mn3X_MS2020,Mn3Ga_2020,AFM_AHE_T2}.

Here we report the Sb doping effect on FeGe and mapping the phase diagram of FeGe$_{1-x}$Sb$_x$ ($0\textless x \textless 0.4$). Using x-ray, transport, magnetic susceptibility and neutron scattering measurements, we characterize the evolution of crystal structure, CDW and magnetic order with Sb doping. Intriguingly, noncollinear AFM structures are found to widely exist in FeGe$_{1-x}$Sb$_x$. The studies on this new series of kagome metals may not only provide opportunities to understand the origin of unconventional CDW and its interplay with magnetic order in FeGe, but also could stimulate future researches on exploring novel topological and correlated phenomena driven by kagome physics.        

\begin{figure*}[htbp]
	\centering
	\includegraphics[width=\textwidth]{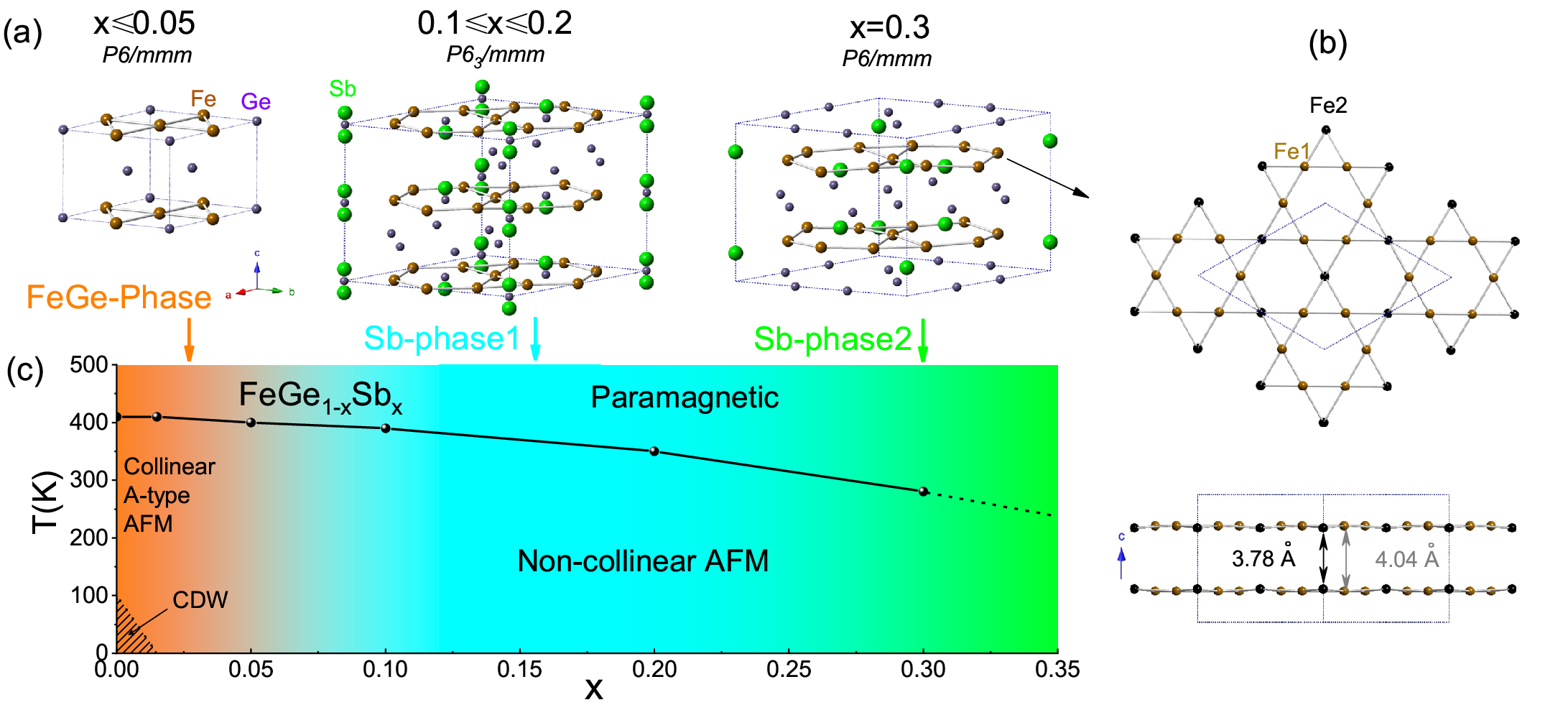}
	\caption {(a) Illustration of the crystal structures of FeGe$_{1-x}$Sb$_x$ at different doping concentration x. The crystal unit cell is marked by blue dotted lines. For x$\textgreater$0.1, some adjacent atoms in extended cells along $c$-axis are also shown for clarity. (b) For x=0.3, The Fe kagome lattice in the $ab$-plane is shown on the top. On the bottom, another view shows that the kagome lattice is slightly distorted along the $c$-axis. (c) Compositional-temperature phase diagram for FeGe$_{1-x}$Sb$_x$.} \label{Fig1}
\end{figure*}

\section{methods}

Polycrystalline FeGe$_{1-x}$Sb$_x$ samples were synthesized by solid state reaction of stoichiometric Fe, Ge and Sb powders at 700$\,^{\circ}\mathrm{C}$ for 4 days, then furnace-cooled to room temperature. The samples with x$\textless$0.3 are characterized by powder x-ray diffraction (XRD) using a Bruker D8 Advance X-ray diffractometer and appear to be phase-pure. For x=0.33, some minor impurity phases including Fe$_3$Ge$_2$ and Sb could be identified (less than 9$\%$). 

Single crystals of FeGe$_{1-x}$Sb$_x$ were grown by the chemical vapour transport method using the synthesized polycrystalline samples similar as previous reports\cite{FeGe}. The obtained crystals are three-dimensional with typical size of 1~mm. The elemental composition of all single crystals were characterized with energy dispersive x-ray spectroscopy (EDS, Oxford X-Max 50). The doping concentration x determined by EDS may have slight deviation from the nominal doping value. For example, the single crystals with nominal x=0.01 are determined to be x=0.015 by EDS. All values of $x$ refer to the EDS values in this manuscript except for polycrystalline samples. The crystal structure of single crystals were all examined by a Bruker D8 VENTURE single-crystal diffractometer using Cu K$_{\alpha}$ radiation and the lattice parameters are determined by refinement.

Magnetization and electrical transport measurements were carried out in Quantum Design MPMS3 and PPMS-14T, respectively. The powder neutron diffraction experiments were carried out on Xingzhi cold neutron triple-axis spectrometer at the China Advanced Research Reactor (CARR)\cite{XingZhi}. About 4-6~g FeGe$_{1-x}$Sb$_x$ powders for each doping were used in the neutron experiments. The incident neutron energy is fixed at 16~meV. The program FullProf Suite package was used in the representational analysis and Rietveld refinement of neutron powder diffraction data\cite{Fullprof}.

\section{Results and discussions}

\begin{figure*}[htbp]
	\centering
	\includegraphics[width=\textwidth]{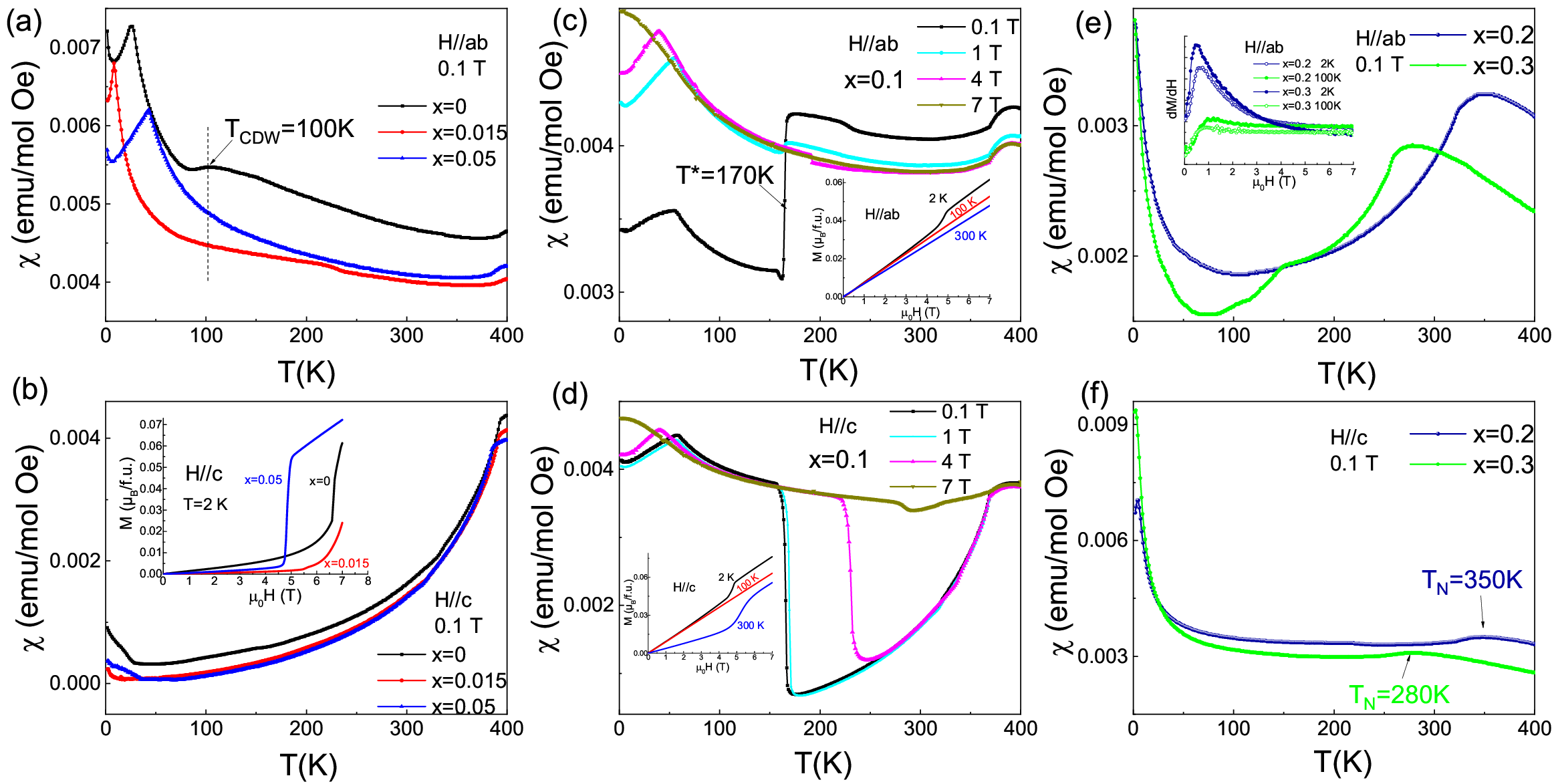}
	\caption {Temperature dependent magnetic susceptibilities for FeGe$_{1-x}$Sb$_x$ single crystals under magnetic field applied parallel to the $ab$-plane and along the $c$-axis. The data of x=0, 0.015 and 0.05 with FeGe-phase are plotted in (a) and (b). For x=0.1 with Sb-phase1, the transition at T$^*$=170~K and its evolution with field can be seen in (c) and (d). The data of x=0.2 and 0.3 are shown in (e) and(f). The insets show the isothermal M(H) curves measured at different field and $dM/dH$ as a function of field is plot in the inset of (e) for a clear view of field-induced magnetic transition.} \label{Fig2}
\end{figure*}

Hexagonal FeGe adopts a CoSn-type crystal structure with alternating stacking of Fe$_3$Ge kagome layer and Ge honeycomb layer. Our XRD analysis on both single crystals and polycrystalline samples reveal that FeGe$_{1-x}$Sb$_x$ maintains the crystal structure of FeGe for x$\leqslant0.05$. However at higher doping level, the results show that Sb does not simply replace Ge and new chemical phases are formed as illustrated in Fig. 1(a). It should be mentioned that the crystal structures of FeGe$_{1-x}$Sb$_x$ with x$\geqslant0.1$ were initially determined by Mills $et~al.$ in an early publication\cite{FeGeSb1} and consistent with our results here. We name the FeGe$_{1-x}$Sb$_x$ with $0.1\leqslant$x$\leqslant0.2$ as Sb-phase1 and that with x=0.3 and 0.33 as Sb-phase2. As shown in Fig. 1(a), the unit cells of new phases are all about six times larger than that of FeGe-phase ($a'$=$\sqrt{3}a$, $c'$=2$c$). The Sb-phase1 adopts a different space group $P6_3/mmm$. The Ge atoms in the honeycomb layer are gradually removed while the Sb atoms form Sb$_2$ pairs whose center of mass lying at the center of hexagons in the Fe$_3$Ge kagome plane. The occupancy of Sb$_2$ pairs is only partial. For Sb-phase2, the structure can be best described using the chemical formula Fe$_3$Ge$_2$Sb. Comparing with FeGe-phase, the Ge honeycomb layer in the Sb-phase2 remains unchanged while the Ge atoms in the Fe$_3$Ge kagome layer are completely replaced by Sb atoms whose positions have an ordered shift along the $c$-axis.

\begin{figure}
	\includegraphics[width=7cm]{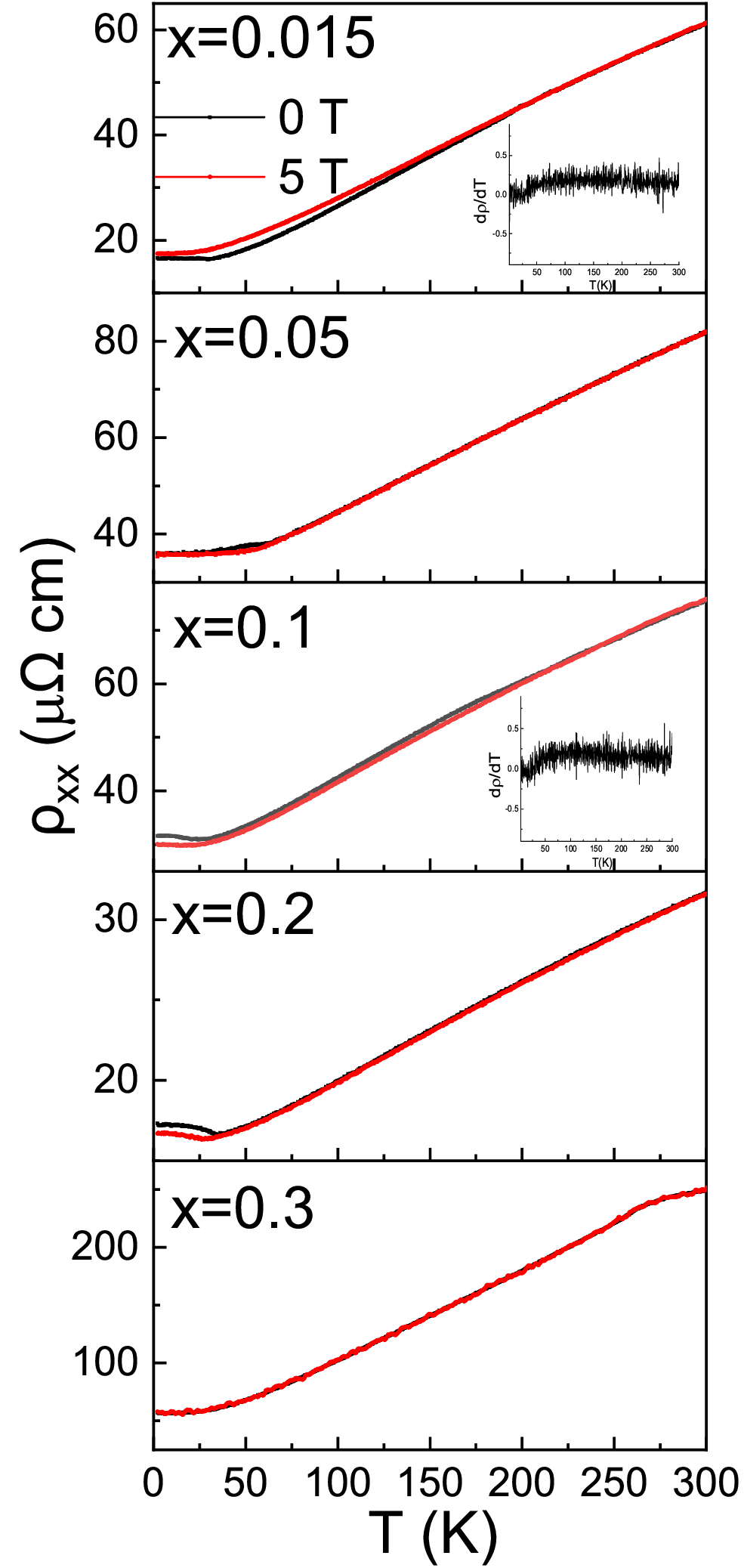}
	\caption {Temperature dependent resistivity of FeGe$_{1-x}$Sb$_x$ single crystals under $\mu_0$H=0 and 5~T. The insets show the $d\rho/dT$ curves for some samples.} \label{Fig3}
\end{figure}

Then we focus on the structural details in the Fe kagome lattice. As shown in Fig. 1(b), different from the FeGe-phase, the Fe ions occupy two inequivalent Wyckoff positions for both Sb-phase1 and Sb-phase2. The nearest Fe1-Fe1 distance between two adjacent kagome layer is larger than the Fe2-Fe2 distance as illustrated in Fig. 1(b). This results in slight distortion of the kagome layer along the $c$-axis comparing with the perfect flat kagome net in FeGe-Phase.

\begin{table}[htbp]
	\caption{Room temperature lattice parameters for FeGe$_{1-x}$Sb$_x$ obtained from XRD.} \label{tab.1}
	\begin{center}
		\begin{tabular}{ccccccccc}
			sample type & x & $a$(\AA) & $c$(\AA)  \\
			\hline
			Single crystal & 0 & 5.003 & 4.055 \\
			Single crystal & 0.015 & 5.031 & 4.055 \\
			Single crystal & 0.05 & 5.063 & 4.056 \\
			Single crystal & 0.1 &
			8.830 & 8.108 \\
			Single crystal & 0.2 & 8.930 & 7.990 \\
			Single crystal & 0.3 & 8.976 & 7.952\\		
			Polycrystalline & 0.1 & 8.769 & 8.037 \\
			
			Polycrystalline & 0.2 & 8.873 & 7.978 \\
			
			Polycrystalline & 0.33 & 8.931 & 7.942 \\
		\end{tabular}
	\end{center}
\end{table}

Fig. 1(c) presents the phase diagram of FeGe$_{1-x}$Sb$_x$ which shows the evolution of different solid phases with doping concentration. The room temperature lattice parameters determined from XRD for different samples are presented in Table I. A general tendency is that with increasing $x$, the $a$-axis lattice constant increases while the $c$-axis constant decreases. The lattice constants of FeGe-phase could be transformed by using $a'$=$\sqrt{3}a$ and $c'$=2$c$ formulas for comparison. It seems that the doping of Sb causes a lattice compression effect along the $c$-axis. As a result, the nearest Fe-Fe distance in one kagome layer increases from 2.54$\AA$ for FeGe to 2.60$\AA$ for Fe$_3$Ge$_2$Sb. The buckled kagome structure of Fe$_3$Ge$_2$Sb is also confirmed by a very recent publication\cite{FeGeSb2}. For the physical properties of FeGe$_{1-x}$Sb$_x$, so far as we know, only that of Fe$_3$Ge$_2$Sb (close to our samples with x=0.3 and 0.33) was reported recently\cite{FeGeSb2}.

FeGe serves as a very rare example for the coexistence of CDW and AFM order. We firstly investigate how these orders would evolve with Sb doping via magnetization measurements. Fig. 2(a) and (b) shows the temperature dependent magnetic susceptibility of single crystals with FeGe-Phase. Consistent with previous report, for the parent compound FeGe, there is a hump at 100~K in the $\chi$(T) curve under H$\parallel$ab due to the development of CDW order\cite{FeGe}. However this feature disappear in x=0.015, indicating a rapid suppression of CDW with Sb doping. In addition, FeGe was reported to have a spin-flop transition under H$\parallel$c. It is found that the transition field shifts from 7~T to about 5~T at 2~K as shown in the inset of Fig. 2(b).

\begin{figure*}[htbp]
	\centering
	\includegraphics[width=\textwidth]{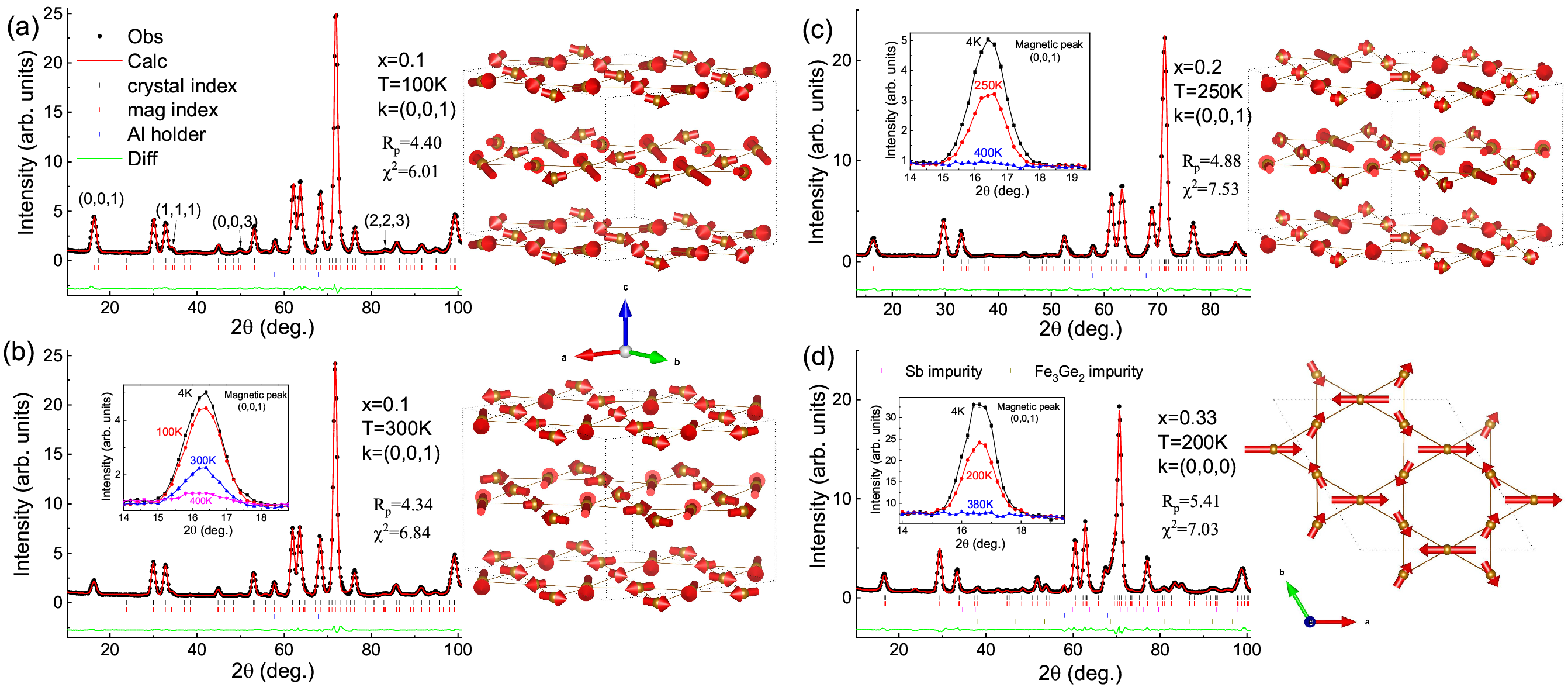}
	\caption {Neutron diffraction patterns, Rietveld refinement results and corresponding magnetic structures of x=0.1, 0.2 and 0.33 at different temperatures are shown respectively. The indices of four magnetic Bragg peaks are labeled in (a). The insets of (b), (c) and (d) show the (0,0,1) magnetic peak at different temperatures. The magnetic unit cell is marked by dotted lines and solid lines indicate the nearest Fe-Fe bond in the $ab$-plane.} \label{Fig4}
\end{figure*}

As revealed from Fig. 2, the AFM transition temperature is gradually suppressed to lower temperature with increasing x. T$_N$ is determined to be 350~K for x=0.2 and 280~K for x=0.3. Another important feature is that, for samples with FeGe-Phase, the susceptibility has a much sharper drop below T$_N$ under H$\parallel$c in contrast with that under H$\parallel$ab. This is a typical feature for antiferromagnet with ordered moment parallel to the $c$-axis. However for samples with Sb-phase1 and Sb-phase2, this feature is reversed. The susceptibility drop is much sharper under H$\parallel$ab for x=0.2 and 0.3 (Fig. 2(e) and (d)), which suggests the magnetic moments tend to lie in the $ab$-plane. This doping induced change of magnetic anisotropy is also confirmed by the following neutron diffraction studies. Besides, a sudden jump of susceptibility occurs at T$^*$=170~K under $\mu_0$H=0.1T and moves to higher temperature with increasing field. It is likely caused by a temperature-driven spin-reorientation transition since magnetic field has strong impact on it. In addition, magnetic field induced spin-flop transitions could be identified for x=0.1 under H$\parallel$ab and H$\parallel$c from the M(H) curves in the insets of Fig. 2(c) and (d). For x=0.2 and 0.3, field-induced spin-flop transitions may exist under H$\parallel$ab as revealed from the dM/dH curves in the inset of Fig. 2(e), while similar features are not observed in M(H) curves for H$\parallel$c within the field limit.

Temperature dependent electrical resistivity measured under $\mu_0$H=0 and 5~T are displayed in Fig. 3. For FeGe with the charge order, a kink occurs at the CDW transition temperature in the $d\rho/dT$ curve as reported previously\cite{FeGe}. However, this feature disappears for x=0.015 as shown in the inset of Fig. 3. For x=0.1, no distinguishable anomaly is identified in the $d\rho/dT$ curve across the possible spin-reorientation transition at T$^*$=170~K. A weak negative magnetoresistance (MR) could be observed below T$^*$ and becomes notable below 30~K. Interestingly, the MR is positive for x=0.015 and nearly zero for x=0.3, negative MR only becomes visible for x=0.1 and 0.2. We speculate the MR behavior might be associate with the magnetic structure, magnetic field may reduce the strong spin scattering caused by the noncollinear AFM structure and results in negative MR. In addition, for both x=0.1 and 0.2, an upturn of resistivity occurs below 30~K which may possibly be due to the disorder induced localization effect. As these samples with Sb-phase1 have significant atomic vacancies in the Ge and Sb sites.   

Next, we present powder neutron diffraction results on x=0.1, 0.2 and 0.33. For all three samples, the most prominent and well defined magnetic Bragg peak is indexed as (0,0,1) as seen from the insets in Fig. 4. According to the basic magnetic neutron scattering rules, if the ordered moments strictly lie parallel to the $c$-axis, then (0,0,1) should have zero intensity contribution from magnetic scattering, which is the case of FeGe at between 400~K and 60~K as seen from previous neutron scattering experiments\cite{FeGe,FeGe_1978,FeGe_1984}. Please be aware that since the $c$-lattice constant is doubled for FeGe$_{1-x}$Sb$_x$ comparing with that of FeGe, the (0,0,1) for x$\geqslant$0.1 should be considered as (0,0,0.5) for FeGe. So the significant magnetic contribution for (0,0,1) clearly means that the ordered moments of FeGe$_{1-x}$Sb$_x$ should have dominant in-plane components. 

For x=0.1 and 0.2 with Sb-phase1, (0,0,1) and other notable magnetic peaks includes (1,1,1), (0,0,3) and (2,2,3) which should be in structural extinction (Fig. 4(a)). This set of magnetic peaks are well defined by a propagation vector $\mathbf{k}$=(0,0,1). Then we employed the BasIreps program to carry out representational analysis\cite{Fullprof}. The result reveals twelve irreducible representations (IR) for Fe1 and six IRs for Fe2 which are compatible with this propagation vector. Each IR describes a possible magnetic model and we find only one IR for both Fe1 and Fe2 could give the best fit of the diffraction data, the fitting with other IRs yields unacceptable $R_P$ and $\chi^2$ factors. The refinement results and corresponding magnetic structure for x=0.1 and x=0.2 are shown in Fig. 4(a), (b) and (c). Apparently, all magnetic structures are noncollinear and the ordered moments at different Fe-sites are ranging from $\sim$1$\mu_B$ to $\sim$4$\mu_B$. For x=0.1 at 100~K, the ordered moments have small components along $c$-axis (less than 0.9$\mu_B$) which makes the AFM structure noncoplanar. While at 300~K, the $c$-axis component becomes negligible small (less than 0.01$\mu_B$) and the orientation of the in-plane components also have some changes. This might explains the spin-reorientation transition at T$^*$=170~K. For x=0.2 at 250~K, the $c$-axis component is zero and the AFM structure is coplanar. The AFM structures at 4~K are similar since the magnetic peaks are the same and only have some intensity changes. The detailed data of the magnetic structures of all samples at different temperatures derived from refinement are recorded as '$.mcif$' files provided in the supplemental materials. 

For x=0.33 with Sb-phase2, the indexed magnetic peaks are similar but they are no longer in structural extinction due to the different crystal symmetry, so a propagation vector $\mathbf{k}$=(0,0,0) is chosen. Similar representation analysis and Rietveld refinement process also reveal that only one IR could best fit the diffraction data. Interestingly, for all the AFM structures in Sb-phase1, the in-plane components of the ordered moments are antiparallel between adjacent layers suggesting an interlayer AFM interaction. However for x=0.33, the in-plane basic vectors of the only IR which could fit the data are parallel with each other for atoms with the same $z$-axis coordinate, which yields an magnetic structure with interlayer ferromagnetic coupling. This spin configuration is illustrated from a view in the $ab$-plane as shown in Fig. 4(d). According to the fitting result, the spins are coplanar and aligned in a 120$^{\circ}$ triangle AFM type. 

We should mention that typically for a complex noncollinear AFM structure, neutron diffraction on single crystals might be essential for an accurate determination of the magnetic structures. Currently the limited size of FeGe$_{1-x}$Sb$_x$ is hindering us to step forward. However, our the powder neutron diffraction results could at least confirm the existence of noncollinear magnetic structure. Actually, all basic vectors of the possible IRs have noncollinear components in the $ab$-plane, therefore a noncollinear AFM structure is inevitable for the propagation vector determined by the indices of magnetic peaks.

Finally, we would like to discuss the above results in two aspects. First of all, since slight Sb doping could remove the CDW order, it may provide opportunities to uncover the origin of CDW in FeGe. The recent angle-resolved photo emission spectroscopy (ARPES) study on FeGe proposes that magnetism induced band-splitting pushes the van Hove singularities to the Fermi level, results in the formation of unconventional charge order\cite{FeGe_ARPES}. So it would be an interesting topic to study how the band structure would be affected by slight Sb doping via ARPES and band calculations, which may provide critical information about the origin of CDW order. 

Secondly, a recent theoretical work has predicted that an evolution from intralayer ferromagnetism to 120$^{\circ}$ AFM and noncoplanar spin orders could be realized in kagome metals by tuning onsite repulsion and flat band fillings\cite{arxiv2023}. FeGe was proposed in the border of these noncollinear AFM orders as shown in the theoretical phase diagram\cite{arxiv2023}. These intriguing unconventional noncollinear AFM orders that are closely related to kagome flat band might be realized in FeGe$_{1-x}$Sb$_x$ as demonstrated in our results, although further theoretical and experimental evidences are needed for a final confirmation. The noncollinear AFM structures are very rare in kagome metals besides Mn$_3$X (X= Ge, Sn, Ga, Ir) material family\cite{Mn3X_MS2020,Mn3Ga_2020,AFM_AHE_T2}. Our results may stimulate future researches on exploring anomalous Hall effect and spin Hall effect in FeGe$_{1-x}$Sb$_x$ which may be induced by noncollinear antiferromagnetism.

\section{Conclusions}

In summary, the physical properties and phase diagram of kagome metals FeGe$_{1-x}$Sb$_x$ are presented. The drastic suppression of CDW order and change of magnetic anisotropy with Sb doping are observed. Neutron diffraction investigations reveal that noncollinear magnetic structures develop in FeGe$_{1-x}$Sb$_x$ with buckled kagome lattice which is substantially different from the magnetic structure in the parent compound FeGe. We argue that this noncollinear antiferromagnetism might be unconventional and closely related to the kagome flat band. FeGe$_{1-x}$Sb$_x$ could become a new material platform to explore novel emergent phenomena related to kagome physics.

\section*{Acknowledgement}
This work was supported by the National Natural Science Foundation of China (No. 12074426, No. 12004426, No. 11227906), the Fundamental Research Funds for the Central Universities, and the Research Funds of Renmin University of China (Grants No. 22XNKJ40), NSAF (Grant No. U2030106) and the Outstanding Innovative Talents Cultivation Funded Programs 2023 of Renmin Univertity of China.

\bibliography{FeGeSb}{}
\end{document}